\documentclass[fleqn,twoside]{article}
\usepackage{espcrc2}
\usepackage{graphicx}

\usepackage[figuresright]{rotating}
  
\newcommand{\ttbs}{\char'134}

\title{Confinement \& Chiral Symmetry Breaking:\\ 
The fundamental problems of hadron physics}

\author{Reinhard Alkofer\address{Institute of Physics, University Graz,
Universit\"atsplatz 5, A-8010 Graz, Austria}}
       
\begin{document}

\begin{abstract}
Some of the difficulties arising when one tries to understand confinement  
as well as dynamical and anomalous chiral symmetry breaking are briefly
reviewed. Criteria to be fulfilled by a successful and complete picture of 
these phenomena are presented, and a few of the suggested explanations are
listed.
\vspace{1pc}
\end{abstract}

\maketitle

\section{Phenomenology versus Theory}
The most intriguing phenomena of hadron physics are confinement  as well as
dynamical and anomalous chiral symmetry breaking. Despite the fact that the 
theory of the Strong Interactions,  Quantum  Chromodynamics (QCD),
is known since decades we still lack a fundamental understanding of the
corresponding physics. 

As a phenomenon confinement is easily described. On one hand, representing the
Strong Interaction by a local Quantum Field Theory ({\it i.e.}~by
QCD) necessitates to introduce fundamental fields
with a new quantum number, namely quarks and gluons with some \ttbs
colour\ttbs. The advantage of this approach is twofold: It provides a
mathematical framework, and it orders the plethora of hadrons into a clearly
arranged pattern. On the other hand, quarks and gluons have never been detected
as particles, {\it i.e.} nobody has ever seen quarks and gluons making a
track in a detector. The confinement hypothesis can therefore be formulated as:
the colour-neutral hadrons, being a kind of bound states of coloured quarks and
gluons, are the only strongly interacting particles, no \ttbs coloured\ttbs \
particles exist. This hypothesis has been extremely successful. The
colour-charge version of ionization does plainly not occur. Even more, the
concept of mutual forces by mutual polarization, the van-der-Waals forces, also
does not have a colour-charge analogue. Thus as a phenomenon confinement
seems to be plain and simple.

As a theoretical concept confinement is astonishingly hard to put into precise
terms. Even the question how to obtain a concise definition of `charge' did
undergo some severe discussions when trying to find a theoretically unequivocal 
definition of confinement. {\it E.g.} the Wilson loop provides an order
parameter only in the absence of fundamental charges,  {\it i.e.}~quarks.
Despite all efforts such an order parameter has not been found in the real world
with light quarks, a satisfactory and detailed description of the underlying
mechanisms of confinement stays elusive. The fact that for charges in higher
representations there are common aspects with the Higgs mechanism
complicates the issue even further.

The situation is not drastically different when addressing dynamical Chiral
Symmetry Breaking ($\chi$SB) and the $U_A(1)$ anomaly. As phenomena they are
clearly identifiable, the first because of the relatively small pion mass and
several patterns in the interaction of pions with themselves and other hadrons.
The latter because of the large $\eta^\prime$ mass.

When it comes to theoretically understanding $\chi$SB we also lack a lot of
basic knowledge. We  know that dynamical $\chi$SB comes along with the dynamical
generation of \ttbs constituent\ttbs quark masses (which, however, depend on the
momentum of the quarks). One may explain dynamical $\chi$SB and the $U_A(1)$
anomaly with two seemingly different approaches. One approach starts by
considering quark zero modes in topologically non-trivial field configurations.
A non-vanishing density of such zero modes in the limit of infinite volume
signals  dynamical $\chi$SB  \cite{Banks:1979yr}. The non-vanishing topological
susceptibility provides the explicit $U_A(1)$ symmetry breaking, see {\it
e.g.\/} \cite{Leutwyler:1992yt} and references therein.

The other approach rests on a supercritical effective interaction between
quarks \cite{Miransky:1985ib,Pennington:1998cj}, 
usually described in a covariant Green's function approach see {\it e.g.\/} 
\cite{Alkofer:2000wg,Fischer:2006ub,Alkofer:2006jf,Aguilar:2008xm,Dudal:2008rm,Fischer:2009tn} 
and references therein. The mass generating
mechanism becomes then similar to the generation of a gap in superconductors.
Especially, if this interaction is infrared divergent the effective coupling
always exceeds the critical one and therefore dynamical $\chi$SB occurs.
What is more astonishing is the fact that a confining-type infrared divergence in
the effective quark-quark interaction results in a non-vanishing $\eta^\prime$
mass \cite{Kogut:1973ab,Alkofer:2008et}. Therefore it may well be possible that
these two so differently appearing approaches are merely two distinct but
correct ways of describing the related physics and aspects thereof.

As we have no commonly accepted complete picture of the strongly interacting
domain of QCD  the relation between confinement on the one hand and dynamical,
resp., anomalous, $\chi$SB on the other hand is not firmly established. However,
there are important hints that quark confinement and $\chi$SB are closely
related. Even beyond the debated question whether the corresponding phase
transition(s) occur(s) at the same temperature (see {\it e.g.\/}
\cite{Bazavov:2009zn} and references therein) an analysis of the so-called dual
quark condensate and dressed Polyakov loops points to such a close relation
\cite{Bilgici:2008qy,Fischer:2009wc,Bilgici:2009tx} 
via linking confinement to spectral
properties of the Dirac operator \cite{Gattringer:2006ci}. Again such a close
relation can be found in the approaches mentioned above: Either when
investigating topologically non-trivial, confining field configurations 
\cite{Di Giacomo:1999fa,Greensite:2003bk,Diakonov:2009jq}  or when studying
the infrared behaviour of QCD Green functions, and hereby especially the
quark-gluon vertex in Landau gauge
\cite{Alkofer:2008tt,Alkofer:2006gz}. But despite all evidence for a deep
connection between confinement and $\chi$SB the situation is
not conclusive yet.

\section{Remarks on Quantum Field Theory}

According to my understanding QCD is a {\bf local} Quantum Field Theory as
expressed in the quote from Haag's book~\cite{Haag:1992hx} in a clear way 
as follows:

{\sl ``The r\^ole of fields is to implement the principle of locality. The
number and the nature of different basic fields needed in the theory is
related to the charge structure, not to the empirical spectrum of particles.''}

To put this understanding in a more precise setting:  I assume validity of 
the Osterwalder-Schrader axioms \cite{Osterwalder:1973dx} except reflection
positivity. This provides a well-defined mathematical framework as described in
refs.\ \cite{Haag:1992hx,Nakanishi:1990qm} and a number of other monographs. It
is important to note that all methods in Quantum Field Theory, including
perturbation theory, lattice field theory, and functional approaches, rely on
this framework. If it were true that QCD is not a local theory more or less all
attempts to understand hadron physics from QCD are questionable. Fortunately, 
the results obtained from QCD provide evidence for the validity of locality.

Gaining an understanding of physics is quite often related to develop intuitive
pictures. In the case of confinement such a picture will be preferentially 
formulated with the help of the fundamental fields, the gluons and quarks. But
these are only valid elements of the theory after gauge-fixing. Of course,
confinement as an observable phenomenon exists without reference to any gauge,
and in different gauges picturing confinement might result in quite different
scenarios. However, this is exactly the point. Everybody will agree that the
hydrogen atom can be described by quantum mechanics independent of the gauge
chosen for electromagnetism. For gaining an understanding of the laws of Quantum
Mechanics, however, it was of utmost importance that the spectrum of the
hydrogen atom can be easiest understood when choosing Coulomb gauge. To gain
knowledge in which gauge confinement will be explained easiest would be a
tremendous step forward. Consequently, fixing the gauge is likely to be 
helpful for an understanding of confinement.

\begin{figure}[htb]
\includegraphics[width=75mm]{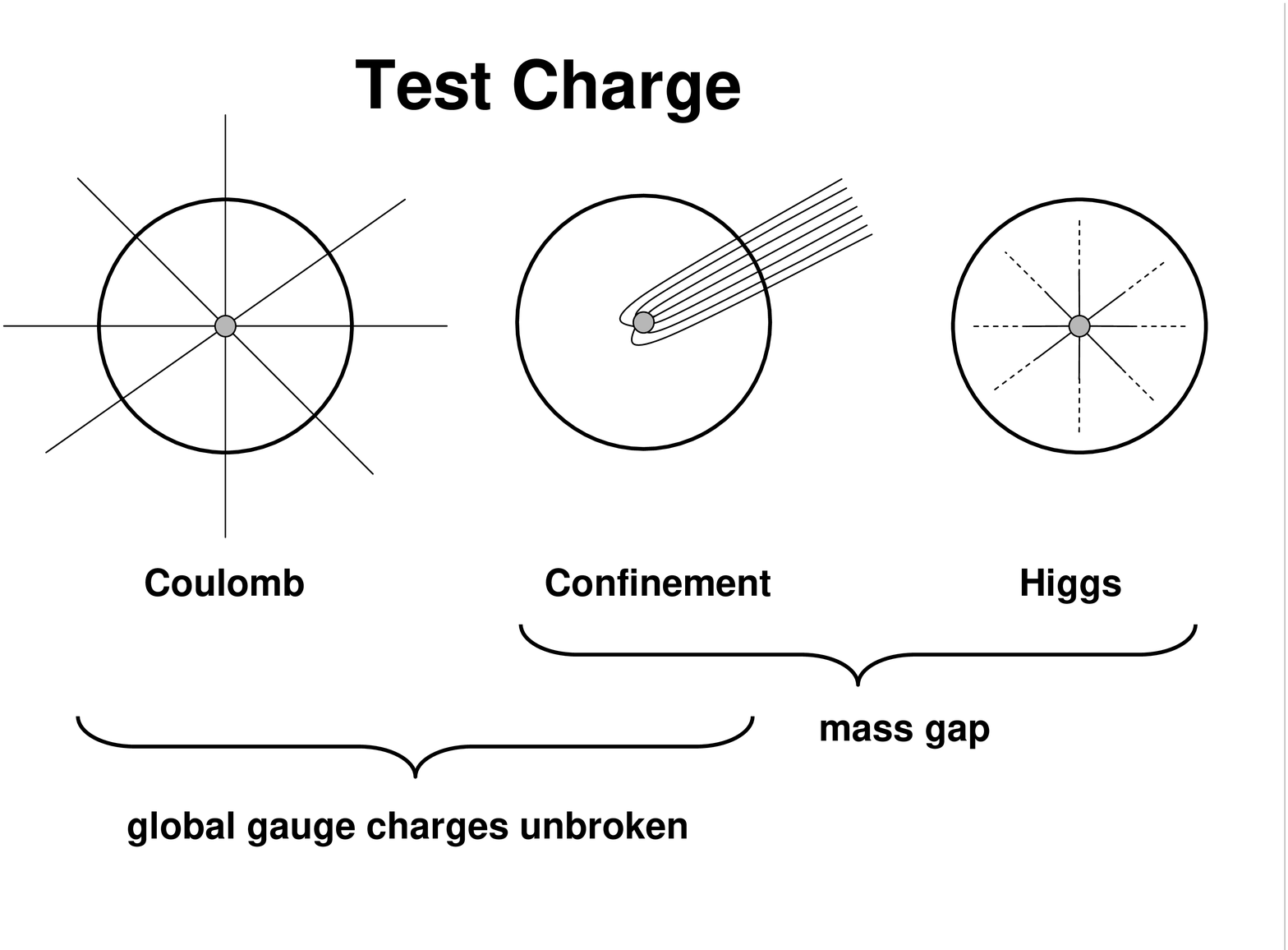}
\caption{A pictorial presentation how the field around a test charge and the
(non-)existence of a mass gap allows to distinguish between the Coulomb,
confinement and Higgs phase \cite{vonSmekal:2008ws}.}
\label{Conf}
\end{figure}

As already mentioned the Wilson loop gives only a clear criterion in the absence
of quarks. So, what are the possiblities for a theoretically sound definition of
confinement? A potential procedure may look like:
\begin{itemize}
\item Construct a colour charge operator,  {\it e.g.} as described in
\cite{Nakanishi:1990qm},
\item demonstrate it to be well-defined (``unbroken charge"), and
\item check for a mass gap in the physical state space.
\end{itemize}
In case one obtains a well-defined charge with unbroken global symmetry 
and a mass gap in the physical state space one has confinement
\cite{vonSmekal:2008ws}. As pictorially
presented\footnote{I thank Lorenz von Smekal for this figure.} 
in fig.~\ref{Conf} 
the unbroken global symmetry without a mass gap provides the
Coulomb-type phase whereas broken global symmetry with mass gap gives the Higgs
phase.

To conclude this section let me emphasize the r\^ole of the 
Becchi-Rouet-Stora--Tyutin (BRST) symmetry in gauge-fixed quantum gauge field
theories. The existence of BRST quartets and the construction of a  BRST
cohomology does not only allow the generalization of the Gupta-Bleuler mechanism
of QED to QCD but also very likely is substantial in constructing the physical
state space. The distinction  between the complete and the positive-definite
state space is hereby absolutely crucial in understanding the mathematical
framework of quantum gauge field theories. An introduction to the subject can be
found in ref.~\cite{Nakanishi:1990qm}, a short summary on how this may relate to
the confinement problem in ref.~\cite{vonSmekal:2000pz}. 

\section{Requirements for an investigation of Confinement}

First, confinement in four-dimensional field theories requires the dynamical 
generation of a physical mass scale. In
presence of such a mass scale, however, the renormalisation group (RG)
equations imply the existence of essential singularities in physical
quantities (such as the $S$-matrix) as functions of the coupling at $g =
0$. This is due to the dependence of the RG invariant confinement scale
on the coupling and the renormalisation scale $\mu$ near the ultraviolet
fixed point as given by 
\begin{eqnarray}
  \Lambda &=& \mu \exp \left( - \int ^g \frac {dg'}{\beta (g')} \right) 
  \nonumber \\
 &\stackrel{g\to 0}{\rightarrow } &\mu  \exp \left( - \frac 1 {2\beta_0g^2}
\right),   
  \label{Lambda}
\end{eqnarray} 
with $\beta_0>0 $.
Therefore a  truely non-perturbative method is needed for the study of
confinement. 

Second, in some scenarios 
confinement is related to severe infrared divergences, {\it i.e.}, divergences
which cannot be removed from physical cross sections by a suitable summation
over degenerate states as in QED.\footnote{See, however, ref.\ 
\cite{Braun:2007bx} which shows that confinement criteria can be fulfilled
without infrared divergences.} In any finite volume these infrared
divergences could be detected only by a careful extrapolation to infinite volume.
Therefore either such an analysis of lattice results and/or an ab initio
continuum approach is needed for an understanding of such confinement scenarios. 

Third, confinement implies the suppression of long-wavelength propagation.
Phrased otherwise, confinement is a true quantum phenomenon. Therefore a purely 
(semi-)classical description is necessarily incomplete, a quantum theoretical
picture is needed for an investigation of confinement.

\section{Criteria for a Confinement picture}

A successful confinement scenario should explain many properties either
deduced from hadron physics or lattice calculations. One of them is
\begin{itemize}
\item  {\em string formation.}
\end{itemize}
There are two distinct sorts of representation dependence of the static quark
potential, 
depending on the static source separation:
\begin{itemize}
\item {\em Casimir Scaling.}  Initially the slope of the linear potential 
$-$ the string tension $-$ is proportional
to the quadratic Casimir of the group representation.
\item {\em N-ality Dependence.}  Asymptotically, the force between charged 
fields in an SU(N) gauge theory depends
only on the so-called ``N-ality" of the group representation, 
given by the number of boxes mod $N$ in the
Young tableau of the representation.
\end{itemize}

Another such property is the 
\begin{itemize}
\item   {\em absence of van-der-Waals forces}
\end{itemize}
as discussed in the introductory section.

Related to the issue of the mathematical framework of the theory is the property
of 
\begin{itemize}
\item  {\em positivity violation}
\end{itemize}
and hereby
\begin{itemize}
\item  {\em the BRST quartet mechanism} for tree-level-positive fields.
\item  {\em antiscreening beyond perturbation theory} as expressed in the 
Oehme--Zimmermann superconvergence relations\footnote{See {e.g.} refs.\ 
 \cite{Oehme:1980ai,Alkofer:2000wg}}.
\end{itemize}

And, last but not least, as a successful theory of confinement is a theory of
Infrared QCD it should include a
description of
\begin{itemize}
\item  {\em dynamical  $\chi$SB}
\end{itemize}
and the 
\begin{itemize}
\item  {\em $U_A(1)$ anomaly}.
\end{itemize}

\section{Candidates for a Confinement picture}

There are many proposals for the confinement mechanism. 
 It is impossible to  provide an exhaustive list in a short
article, so I will cite only  those proposals which according to my opinion 
seem best supported by existing numerical studies or other
arguments.\footnote{Some of these arguments are briefly reviewed in ref.\
\cite{Alkofer:2006fu}.}

A line of thought is that the QCD functional integral is dominated by some
special class of field configurations which cause the expectation value of a
large Wilson loop to fall off  exponentially with the minimal area of the loop.
The leading candidates for  these special configurations are magnetic monopoles
\cite{Di Giacomo:1999fa}, dyons /calorons  \cite{Diakonov:2009jq,Kraan:1998pm}
or center vortices \cite{Greensite:2003bk}, although other objects  
have been suggested.

A different approach is based on the special properties of quantum fields in
Coulomb gauge, and hereby the existence of the gauge-fixing ambiguity, the
Gribov problem, and the existence of a Gribov horizon plays a special role
\cite{Gribov:1977wm,Zwanziger:1998ez}.

Another idea is, preferentially in Landau gauge,
 to solve non-perturbatively for quark and gluon
propagators and vertex functions, analytically by an infrared expansion of the
complete set of Schwinger-Dyson and Exact Renormalization Group equations, and
numerically by solving a truncated set of these equations, see {\it e.g.}
\cite{vonSmekal:1998is,Pawlowski:2003hq,Alkofer:2004it,Fischer:2008uz} 
and references therein.

 Finally, there is a fascinating relationship between gauge theory in $D=4$
dimensions and string theory quantized in a special ten-dimension background
geometry known as anti-DeSitter space.  This is the AdS-CFT correspondence,
see refs.\  \cite{Maldacena:1998im,Polchinski:2001tt} and many others.

It has turned out that a number of these suggestions are related in interesting
ways:  monopole wordlines are found
to lie on center   vortex worldsheets, and center vortex worldsheets appear to
be crucial in some ways to the confinement scenario in Coulomb gauge.   Both
Coulomb and Landau gauge investigations emphasize the importance of the
Faddeev-Popov operator, and the infrared properties of the ghost propagator.   

\section{Outlook}

In this contribution to a lively on-going discussion I tried to describe what
are the difficulties encountered in the endeavour of studying infrared QCD. It
is striking that after decades of effort we do not understand how the Strong
Interaction really works at long distances. Nevertheless, there \emph{has} been
appreciable progress in this subject. Step by step we uncover surprising details
about confinement and dynamical, resp., anomalous, chiral symmetry breaking.

Between the existing approaches there are not yet understood relations. Although
many details are still missing these relations make plain that the different
confinement pictures are definitely not mutually exclusive. Maybe we will learn
that a non-trivial merger of all these scenarios of Infrared QCD will
eventually  fulfill all the criteria required for a consistent and convincing
description.

Even if this will constitute the major breakthrough
 for theory one should keep in mind
that even then there is still a tough challenge left: Find an experimentally
accessible hadron observable to verify/falsify the presented picture of
confinement and chiral symmetry breaking.

\section*{Acknowledgement} 

I thank the organizers of this Winter School, my  colleagues Christof
Gattringer, Leonid Glozman, Christian Lang, Heimo Latal, and Leopold Mathelitsch
for inviting me, and especially for all their efforts to make this outstanding
school possible. My knowledge about this subject would have not been possible
without the discussions I enjoyed with many outstanding colleagues. I am
grateful to all of them, but amongst them Christian Fischer, Felipe Llanes
Estrada, Jeff Greensite, Axel Maas, 
Jan Pawlowski,
Lorenz von Smekal, and Dan Zwanziger deserve
a special mentioning. Last but not least,
I thank Christian Fischer and Axel Maas for a critical reading of the
manuscript.

\end{document}